\documentclass[preprint,
 amssymb, amsmath,%
 aip,jcp,
groupedaddress,%
]{revtex4-1}

\usepackage{docs}%
\usepackage{bm}%
\usepackage[colorlinks=true,linkcolor=blue,citecolor=blue,urlcolor=blue]{hyperref}%
\usepackage{graphicx}
\usepackage{subcaption}
\usepackage{braket}
\usepackage{color,soul}%

\expandafter\ifx\csname package@font\endcsname\relax\else
 \expandafter\expandafter
 \expandafter\usepackage
 \expandafter\expandafter
 \expandafter{\csname package@font\endcsname}%
\fi
\begin{document}

\title{Multi-time Formulation of Matsubara Dynamics}%

\author{Kenneth A. Jung}%
\email{kenneth.jung@yale.edu}

\author{Pablo E. Videla}%

\author{Victor S. Batista}%
\email{victor.batista@yale.edu}

\affiliation{Department of Chemistry, Yale University, P.O. Box 208107, New Haven, CT 06520-8107}

\date{\today}

\begin{abstract}

Matsubara dynamics has recently emerged as the most general form of a quantum-Boltzmann-conserving classical dynamics theory for the calculation of single-time correlation functions. Here, we present a generalization of Matsubara dynamics for the evaluation of multi-time correlation functions. We show that the Matsubara approximation can also be used to approximate the two-time symmetrized double Kubo transformed correlation function. By a straightforward
extension of these ideas to the multi-time realm, a multi-time Matsubara dynamics approximation can be obtained for the multi-time fully symmetrized Kubo transformed correlation function. Although not a practical method, due to the presence of a phase-term, this multi-time formulation of Matsubara dynamics represents a benchmark theory for future development of Boltzmann
preserving semi-classical approximations to general higher order multi-time correlation functions.

\end{abstract}

\maketitle

\section{Introduction}
\label{sec:intro}

It is undebatable that quantum thermal time correlation functions (TCFs) play a central role in the description of dynamical properties of chemical systems. \cite{Chandler_Book,Nitzan_Book,McQuarrie_Book}
This has in large been driven by linear response theory\cite{Kubo1957}, which connects linear absorption spectroscopy, diffusion coefficients and reaction rates constants 
with single-time correlation functions. 
Although sometimes a complete classical description of these properties suffices, there are plenty of examples were nuclear quantum effects (NQE), such as zero-point energy fluctuations
and tunneling, play crucial roles and modulate the dynamical behavior of the system.\cite{Kuharski1984,Wallqvist1985,Berne1986,Stern2001,Chen2003,Miller2005_1,Morrone2008,
Paesani2009,Paesani2009_1,Ceriotti2013,Ceriotti2016_1} However, despite the great advances in recent years of algorithms for the exact quantum mechanical propagation of small systems comprising few particles,\cite{Wang2015ML,Greene2017,Richings2018} 
the exact full quantum mechanical calculations of TCFs for condensed phases systems involving many degrees of freedom is still impractical.
Therefore, there is great interest in the
development of reliable approximate methods based on classical molecular dynamics that retain the quantum nature of the Boltzmann distribution. 

Over the past three decades significant progress has been made in this direction, with the development of approximate classical-like methodologies that to some extent include quantum statistics, \cite{Cao1994-2,Cao1994-4,Wang1998,Jang1999,
Craig2004,Habershon2013,Rossi2014,liu2014path,Smith2015}
providing efficient and robust
ways of including NQE into dynamical properties like vibrational spectra, diffusion coefficients and
reaction rates constants for a variety of condensed phases systems.\cite{Batista1999,Liu2009,Liu2011,Habershon2008,
Paesani2009,Rossi2014,Medders2015,Medders2016,
Smith2015A,Liu2016,Videla2018}
Very recently, a new
approximation known as Matsubara dynamics
\cite{Hele2015Mats} was derived and 
demonstrated to give the most consistent way of obtaining classical dynamics from quantum dynamics while preserving the quantum Boltzmann statistics. Although not a practical methodology, due to the presence of a phase factor in the quantum distribution that gives rise to a sign problem, Matsubara dynamics represents a benchmark theory to develop  and to rationalize approximate methods. 
Upon performing additional approximations\cite{Hele2015,Willatt2018} to Matsubara dynamics it is possible to obtain previous heuristic methodologies such as centroid molecular dynamics (CMD)\cite{Cao1994-2,Cao1994-4,Jang1999}, ring-polymer molecular dynamics (RPMD)\cite{Craig2004,Habershon2013} and the planetary model\cite{Smith2015,Smith2015A} (also known as the Feynman-Kleinert Quasi Classical Wigner method) but it's true potential comes from it's ability to yield new approximations.\cite{Trenins2018}  
The theory of Matsubara dynamics for the evaluation of single-time correlation functions provides a step towards the correct theoretical description of the combination of classical dynamics and quantum Boltzmann statistics.

However, not all experimental observables can be related to single-time correlations functions. 
Multi-time correlations function involving more than one time variable are of great importance in chemical physics since there are implicated in the description of non-linear spectroscopy\cite{Mukamel_Book,Cho_Book} and non-linear chemical kinetics.\cite{Mukamel2014}  
Very recently we have presented a basic theory\cite{Jung2018} 
that relates two-time double Kubo transformed correlation\cite{Reichman2000} functions with the second order response function 
as a practical approach to include NQE into the simulation of 2D Raman\cite{Ito2015} and 2D Terahertz-Raman spectroscopy.\cite{Savolainen2013}
Obtaining quantum-Boltzmann-conserving classical approximations for the evaluation of multi-time correlation functions is of great interest. Here, we present 
the extension of Matsubara dynamics to the multi-time realm, providing a general quantum Boltzmann preserving classical approximation to multi-time symmetrized Kubo transformed correlation functions.

This paper should be viewed as an extension of the work of Hele et al.\cite{Hele2015Mats} and is organized as follows: we first review the derivation of the single-time Matsubara dynamics approximation\cite{Hele2015Mats} in Sec. \ref{sec:Mats} using a slightly different, albeit completely equivalent, notation that makes use of the Janus operator\cite{Littlejohn1986} along with the Wigner-Moyal series. In Sec. \ref{sec:Two-Time Matsubara Dynamics} we present the two-time Matsubara dynamics approximation.
Sec. \ref{sec:multi-dimensional} presents the generalization of the Matsubara approximation to multi-times. Final remarks and future applications are discussed in Sec. \ref{sec:Conc}.

\section{Single-Time Matsubara Dynamics}
\label{sec:Mats}

To facilitate the derivation of the multi-time Matsubara dynamics approximation, we first review the formulation of single-time Matsubara dynamics.\cite{Hele2015Mats}
This allows us to present the notation that will be used in the paper and to focus on the critical steps of the derivation that will be important for the multi-time generalization.
This section closely follows the derivation presented in Ref. \citenum{Hele2015Mats} and the references therein. The reader is referred to these for further details.

The starting point for deriving Matsubara dynamics is to obtain a path integral discretization of the Kubo transformed single-time correlation function defined by\cite{Kubo1957}
\begin{equation}
K_{AB}(t) = \frac{1}{Z\beta} \int^{\beta}_{0} d\lambda \: Tr\left[e^{-(\beta-\lambda)\hat{H}} \hat{A} e^{-\lambda\hat{H}}e^{i\hat{H}t/\hbar}\hat{B} e^{-i\hat{H}t/\hbar}\right],
\label{eq:singleKubo}
\end{equation}
where $\beta=1/k_BT$ is the inverse temperature and $Z$ is the partition function defined as 
\begin{equation}
Z = Tr\left[e^{-\beta\hat{H}} \right].
\end{equation}
For clarity of presentation, we consider a one-dimensional system with a Hamiltonian of the form $\hat{H}=\hat{p}^2/2m + V(\hat{x})$, with the extension to multidimensional systems being straightforward.\citep{Hele2015Mats} Also, to further simplify the derivation, $\hat{A}=A(\hat{x})$ and $\hat{B}=B(\hat{x})$ are assumed to be position-dependent operators, although similar expressions can be obtained for operators that only depend on the momentum operator.

Discretizing the integral over lambda into $N$ terms, and inserting $N-1$ identities of the form 
\begin{equation}
\hat{1} = e^{i\hat{H}t/\hbar}e^{-i\hat{H}t/\hbar},
\label{eq:TimeI}
\end{equation}
Eq. (\ref{eq:singleKubo}) can be recast as:\cite{Shi2003}
\begin{eqnarray}
K^{[N]}_{AB}(t)
& = & \frac{1}{Z_NN} \sum^{N}_{k=1} Tr \bigg[ \bigg(e^{-\beta_N\hat{H}}e^{i\hat{H}t/\hbar}e^{-i\hat{H}t/\hbar}\bigg)^{N-k-1}
e^{-\beta_N\hat{H}} \hat{A} e^{i\hat{H}t/\hbar}e^{-i\hat{H}t/\hbar} \nonumber \\
& &\times\bigg(e^{-\beta_N\hat{H}}e^{i\hat{H}t/\hbar}e^{-i\hat{H}t/\hbar}\bigg)^{k-1}e^{-\beta_N\hat{H}} 
  e^{i\hat{H}t/\hbar}\hat{B} e^{-i\hat{H}t/\hbar} \bigg] 
\label{eq:SuperDiscKubo}
\end{eqnarray}
where 
\begin{equation}
Z_N= Tr\left[\bigg(e^{-\beta_N\hat{H}}\bigg)^N\right] ,
\end{equation}
and $\beta_N=\beta/N$. 
Note that the symmetric structure of Eq. (\ref{eq:SuperDiscKubo}) allows for 
the interpretation of the trace in terms of a repeating block structure of the form $\hat{J}^{(1)}\equiv e^{-\beta_N\hat{H}}e^{i\hat{H}t/\hbar}e^{-i\hat{H}t/\hbar}$, with the operator $\hat{A}$ evaluated inside a particular block depending on the value of the sum index $k$ and the operator $\hat{B}$ evaluated at the last block.

\begin{figure}[t]
\begin{subfigure}[l]{0.45\textwidth}
	\centering
	\includegraphics[width=0.75\textwidth]{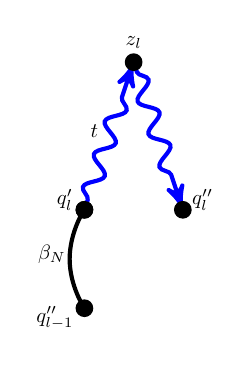} 
\end{subfigure}
\begin{subfigure}[r]{0.45\textwidth}
	\centering
	\includegraphics[width=\textwidth]{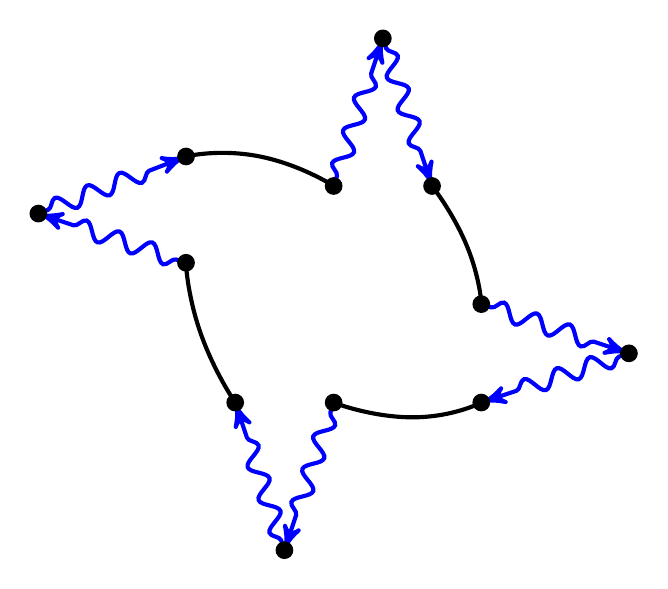}
\end{subfigure}	
\caption{Left: Schematic diagram of the path integral discretization of the block structure $\hat{J}^{(1)}$ (Eq. (\ref{eq:building_1})) used to obtained the generalized Kubo transform in Eq. (\ref{eq:GenKubo}). Blue wavy lines represent the $t$ time propagation. Black lines show the $\beta_N$ imaginary time propagation. Right: Schematic diagram of the  structure of the Kubo transform time correlation function (Eq. (\ref{eq:GenKubo}))
 for $N=4$.}
\protect\label{fig:1}
\end{figure}

A path integral representation of Eq. (\ref{eq:SuperDiscKubo}) can be obtained by inserting identities inside the building blocks $\hat{J}^{(1)}$ in the form (see the schematic representation in Fig. \ref{fig:1})
\begin{eqnarray}
\hat{J}^{(1)}&=&\int dq'_l\int dq''_l\int dz_l \ 
e^{-\beta_N\hat{H}}\ket{q'_l} \bra{q'_l}
e^{i\hat{H}t/\hbar}\ket{z_l} \bra{z_l}
e^{-i\hat{H}t/\hbar}\ket{q''_l} \bra{q''_l},
\label{eq:building_1}
\end{eqnarray}
to yield
\begin{eqnarray}
K_{AB}^{[N]}(t) & = & \frac{1}{Z_N}\int d\boldsymbol{q'}\int d\boldsymbol{q''}\int d\boldsymbol{z}A(\boldsymbol{q'})B(z_N)\prod_{l=1}^{N}\langle q''_{l-1}|e^{-\beta_{N}\hat{H}}\left|q'_{l}\right\rangle\nonumber \\
 &  &  \times\left\langle q'_{l}\right|e^{i\hat{H}t/\hbar}|z_{l}\rangle
\langle z_{l}|e^{-i\hat{H}t/\hbar}\left|q''_{l}\right\rangle,
\label{eq:GenKubo1}
\end{eqnarray}
where 
\begin{equation}
O(\boldsymbol{x})=\frac{1}{N}\sum^{N}_{k=1}O(x_k),
\end{equation}
and $\int d\boldsymbol{x}=\int dx_1\int dx_2\cdots\int dx_N$ (with $x_{0}=x_N$). Note, however, that due to the cyclic structure of 
the path integral representation, the operator $\hat{B}$ can be averaged over all $\boldsymbol{z}$ coordinates to give an even more 
symmetric form of the Kubo transform
\begin{eqnarray}
K_{AB}^{[N]}(t) & = & \frac{1}{Z_N}\int d\boldsymbol{q'}\int d\boldsymbol{q''}\int d\boldsymbol{z}A(\boldsymbol{q'})B(\boldsymbol{z})\prod_{l=1}^{N}\langle q''_{l-1}|e^{-\beta_{N}\hat{H}}\left|q'_{l}\right\rangle\nonumber \\
 &  &  \times\left\langle q'_{l}\right|e^{i\hat{H}t/\hbar}|z_{l}\rangle
\langle z_{l}|e^{-i\hat{H}t/\hbar}\left|q''_{l}\right\rangle.
\label{eq:GenKubo}
\end{eqnarray}
This expression, whose schematic representation is presented in  Fig. \ref{fig:1} for $N=4$, represents an exact path integral representation of Eq. (\ref{eq:singleKubo}) in the limit $N\to\infty$ and emphasizes the symmetry with respect to cyclic permutations of the coordinates of the path integral.\cite{Hele2015Mats,Hele2017}

In order to make the Matsubara approximation, it is necessary to express Eq. (\ref{eq:GenKubo}) in terms of a phase space average.\cite{Wigner1932} To this end, making a change of variables on the Cartesian $\boldsymbol{q'}$ and $\boldsymbol{q''}$ variables to sum/difference coordinates
 \begin{equation}
q_l = \frac{q'_l+q''_l}{2}
\label{eq:SumDiff1}
\end{equation}
\begin{equation}
 \label{eq:SumDiff2}
\Delta_l = q'_l-q''_l
\end{equation}
for each value $l=1,...,N$,
allows us to re-express Eq. (\ref{eq:GenKubo}) as
\begin{equation}
K_{AB}^{[N]}(t) = \int d\boldsymbol{q}\int d\boldsymbol{\Delta}\int d\boldsymbol{z}
A\big(\boldsymbol{q}+\boldsymbol{\Delta}/2\big)
B(\boldsymbol{z})\rho(\boldsymbol{q},\boldsymbol{\ensuremath{\Delta}};\beta)G(\boldsymbol{q},\boldsymbol{\ensuremath{\Delta}},\boldsymbol{z};t),
\label{eq:GenKubo2}
\end{equation}
where we have defined
\begin{equation}
\label{eq:rho}
\rho(\boldsymbol{q},\boldsymbol{\ensuremath{\Delta}};\beta) = \frac{1}{Z_N}\prod_{l=1}^{N}\left\langle q_{l-1}-\frac{\Delta_{l-1}}{2}\right|e^{-\beta_{N}\hat{H}}\left|q_{l}+\frac{\Delta_{l}}{2}\right\rangle,
\end{equation}
and
\begin{equation}
G(\boldsymbol{q},\boldsymbol{\ensuremath{\Delta}},\boldsymbol{z};t) =\prod_{l=1}^{N} \left\langle q_{l}+\frac{\Delta_{l}}{2}\right|e^{i\hat{H}t/\hbar}\Bigl|z_{l}\Bigr\rangle
\Bigl\langle z_{l}\Bigr|e^{-i\hat{H}t/\hbar}\left|q_{l}-\frac{\Delta_{l}}{2}\right\rangle.
\end{equation}
Eq. (\ref{eq:GenKubo2}) is known in the literature as the Generalized Kubo Transformed correlation function\cite{Jang2016,Hele2015Mats}. We anticipate that obtaining an expression of
this form for multi-time correlation functions would be a key step in the multi-time generalization of Matsubara dynamics (see Secs. \ref{sec:Two-Time Matsubara Dynamics} and \ref{sec:multi-dimensional})

Eq. (\ref{eq:GenKubo2}) can be recast as a phase space average by inserting identities of the form
\begin{eqnarray}
\label{eq:Dirac_delta}
1=\int d\Delta'_l \ \delta(\Delta_l+\Delta'_l)=(2\pi\hbar)^{-1}\int d\Delta'_l \int dp_l \ e^{ip_l(\Delta_l+\Delta_l')/\hbar}
\end{eqnarray} 
for each $l=1,...,N$, to obtain \begin{equation}
K_{AB}^{[N]}(t)=\frac{1}{(2\pi\hbar)^{N}}\int d\boldsymbol{q}\int d\boldsymbol{p}\left[e^{-\beta\hat{H}}\hat{A}\right]_{\overline{N}}(\boldsymbol{q},\boldsymbol{p})\left[\hat{B}(t)\right]_{N}(\boldsymbol{q},\boldsymbol{p}),\label{eq:WignerKubo}
\end{equation}
where
\begin{equation}
\left[e^{-\beta\hat{H}}\hat{A}\right]_{\overline{N}}(\boldsymbol{q},\boldsymbol{p})=\frac{1}{Z_N}\int d\boldsymbol{\ensuremath{\Delta}}A\big(\boldsymbol{q}+\boldsymbol{\Delta}/2\big)\prod_{l=1}^{N}\left\langle q_{l-1}-\frac{\Delta_{l-1}}{2}\right|e^{-\beta_{N}\hat{H}}\left|q_{l}+\frac{\Delta_{l}}{2}\right\rangle e^{i\Delta_{l}p_{l}/\hbar},\label{eq:BadWigner}
\end{equation}
and
\begin{eqnarray}
\left[\hat{B}(t)\right]_{N}(\boldsymbol{q},\boldsymbol{p}) & = & \int d\boldsymbol{\ensuremath{\Delta}}\int d\boldsymbol{z}B(\boldsymbol{z})\prod_{l=1}^{N}\left\langle q_{l}-\frac{\Delta_{l}}{2}\right|e^{i\hat{H}t/\hbar}\Bigl|z_{l}\Bigr\rangle\nonumber \\
 &  & \times\Bigl\langle z_{l}\Bigr|e^{-i\hat{H}t/\hbar}\left|q_{l}+\frac{\Delta_{l}}{2}\right\rangle
 e^{i\Delta_{l}p_{l}/\hbar}.\label{eq:GoodWigner}
\end{eqnarray}
The generalized Wigner transform\cite{Wigner1932} in Eq. (\ref{eq:BadWigner}) contains a complex
structure that couples different variables
together whereas the generalized Wigner
transform in Eq. (\ref{eq:GoodWigner}) is just a sum of one-dimensional Wigner
transformed products. Note that since $\hat{B}$ is just a function
of $\hat{x}$ it follows that at
time $t=0$
\begin{equation}
\left[\hat{B}(t=0)\right]_{N}(\boldsymbol{q},\boldsymbol{p})=B(\boldsymbol{q}).
\end{equation}

To complete the phase space representation of the Kubo transform it is useful to recast Eq. (\ref{eq:WignerKubo}) in terms of the quantum Liouvillian instead of the Hamiltonian.
By noting that \cite{Hillery1984,Hele2017}
\begin{eqnarray}
\frac{d}{dt}\left[\hat{B}(t)\right]_{N}(\boldsymbol{q},\boldsymbol{p}) & = & \hat{L}_N\left[\hat{B}(t)\right]_{N}(\boldsymbol{q},\boldsymbol{p}),
\label{eq:Moyal1}
\end{eqnarray}
one can formally write the 
exact correlation function in Eq. (\ref{eq:WignerKubo}) as
\begin{equation}
K_{AB}^{[N]}(t)=\frac{1}{(2\pi\hbar)^{N}}\int d\boldsymbol{q}\int d\boldsymbol{p}\left[e^{-\beta\hat{H}}\hat{A}\right]_{\overline{N}}(\boldsymbol{q},\boldsymbol{p})e^{\hat{L}_{N}t}B(\boldsymbol{q}).
\label{eq:PSKubo}
\end{equation}
In Eqs. (\ref{eq:Moyal1}) and (\ref{eq:PSKubo}) $\hat{L}_N$ represents the Moyal expansion of the quantum Liouvillian of the $N$
blocks as defined by\cite{Hillery1984,Groenewold1946,Moyal1949} 
\begin{equation}
\hat{L}_{N}=\frac{2}{\hbar}\sum_{l=1}^{N}\left[ \frac{p_l^2}{2m} + V(q_l) \right]\sin \bigg( \frac{\hbar}{2} \hat{\Lambda}_l \bigg)
\label{eq:Wigner Moyal}
\end{equation}
where
\begin{equation}
\hat{\Lambda}_l= \frac{\overleftarrow{\partial}}{\partial p_l}\frac{\overrightarrow{\partial}}{\partial q_l} - \frac{\overleftarrow{\partial}}{\partial q_l}\frac{\overrightarrow{\partial}}{\partial p_l},
\label{eq:Janusop}
\end{equation}
is known as the Janus operator\cite{Littlejohn1986} and the arrows indicate the direction in which the differential operators are applied, i.e to the left or right.
In what follows, it will be convenient to 
rewrite the Liouvillian more compactly as
\begin{equation}
\hat{L}_{N}=\frac{2}{\hbar}\left[ \frac{\boldsymbol{p}^2}{2m} + U_N(\boldsymbol{q}) \right]\sin \bigg( \frac{\hbar}{2} \hat{\Lambda}_N \bigg),
\label{eq:Wigner Moyal2}
\end{equation}
where
\begin{equation}
U_{N}(\boldsymbol{q})=\sum_{l=1}^{N}V(q_{l}),
\end{equation}
\begin{equation}
\hat{\Lambda}_N=\mathbf{\overleftarrow{\nabla}}_{\boldsymbol{p}}\cdot\mathbf{\overrightarrow{\nabla}}_{\boldsymbol{q}} - \mathbf{\overleftarrow{\nabla}}_{\boldsymbol{q}}\cdot\mathbf{\overrightarrow{\nabla}}_{\boldsymbol{p}}
\label{eq:Janusqp}
\end{equation}
 and 
\begin{equation}
\mathbf{\nabla}_{\boldsymbol{q}}^{T}=\left(\frac{\partial}{\partial q_{1}},\frac{\partial}{\partial q_{2}},\ldots,\frac{\partial}{\partial q_{N}}\right).
\end{equation}

The advantage of having the Kubo transformed correlation function expressed as a phase space average in ring polymer coordinates, namely Eq. (\ref{eq:PSKubo}), is that now it is possible to make a coordinate transformation to the normal modes describing the centroid and the fluctuations of the free ring polymer.\cite{Richardson2009,Markland2008}
The normal transformation is defined by the matrix $\underline{\underline{\rm T}}$ with elements
\begin{equation}
T_{lk}=\begin{cases}
\sqrt{\frac{1}{N}} & k=0\\
\sqrt{\frac{2}{N}}\sin(2\pi lk/N) & 0<k\leq\bar{N}\\
\sqrt{\frac{2}{N}}\cos(2\pi lk/N) & -\bar{N}\leq k<0,
\end{cases}\label{eq:NMmatrix}
\end{equation}
where $\bar{N}=(N-1)/2$ is chosen to be odd for convenience (even N leads to the same result\cite{Hele2015Mats}). The normal mode coordinates (denoted as $\boldsymbol{Q}$ and $\boldsymbol{P}$)
are defined by the relations
\begin{eqnarray}
\label{eq:NM_Q}
\boldsymbol{q} = \sqrt{N}\underline{\underline{\rm T}}\boldsymbol{Q}
\end{eqnarray}
and
\begin{eqnarray}
\label{eq:NM_P}
\boldsymbol{p} = \sqrt{N}\underline{\underline{\rm T}}\boldsymbol{P},
\end{eqnarray}
where we have included an extra $\sqrt{N}$ factor to ensure that the
$\boldsymbol{Q}$ converge in the limit $N\to\infty$, giving the centroid for $Q_0$.
The square roots of the eigenvalues of the matrix defined in Eq. (\ref{eq:NMmatrix})
are given by
\begin{equation}
\tilde{\omega}_{k}=\frac{2\sin(k\pi/N)}{\beta_{N}\hbar}.\label{eq:Nfreqs}
\end{equation}
In these new coordinates Eq. (\ref{eq:PSKubo}) takes the form
\begin{equation}
K_{AB}^{[N]}(t)=\frac{1}{(2\pi\hbar)^{N}}\int d\boldsymbol{Q}\int d\boldsymbol{P}\left[e^{-\beta\hat{H}}\hat{A}\right]_{\overline{N}}(\boldsymbol{Q},\boldsymbol{P})e^{\hat{L}_{N}t}B(\boldsymbol{Q}),
\label{eq:Kubo_NM}
\end{equation}
where $\int d\boldsymbol{X}=\int dX_{-\bar{N}}\int dX_{-\bar{N}+1 }\cdots\int dX_{\bar{N}-1}\int dX_{\bar{N}}$ and
where the Liouvillian in these new coordinates is given by
\begin{equation}
\hat{L}_{N}=\frac{2N}{\hbar}\left[ \frac{\boldsymbol{P}^2}{2m} + \frac{U_N(\boldsymbol{Q})}{N} \right]\sin \bigg( \frac{\hbar}{2N} \hat{\Lambda}_N \bigg),
\label{eq:Wigner MoyalNM}
\end{equation}
with the Janus operator now defined as
\begin{equation}
\hat{\Lambda}_N=\overleftarrow{\nabla}_{\boldsymbol{P}}\cdot\overrightarrow{\nabla}_{\boldsymbol{Q}}-\overleftarrow{\nabla}_{\boldsymbol{Q}}\cdot\overrightarrow{\nabla}_{\boldsymbol{P}}
\label{eq:JanusNM}
\end{equation}
Following the notation of T. J. Hele et al.,\cite{Hele2015Mats} in the above equations and for what follows it is to be understood that all instances of
$\boldsymbol{Q}$ ($\boldsymbol{P}$) inside a function should be interpreted as $\sqrt{N}\underline{\underline{T}}\boldsymbol{Q}$ ($\sqrt{N}\underline{\underline{T}}\boldsymbol{P}$), i.e. the coordinate transformations defined in  Eqs. (\ref{eq:NM_Q}) and (\ref{eq:NM_P}).

The $M$ lowest frequencies of Eq. (\ref{eq:Nfreqs}) in the limit
as $N\to\infty$ are known in thermal field theories as the Matsubara
modes of distinguishable particles.\cite{Matsubara1955} An explicit form
for these frequencies exists,
\begin{equation}
\lim_{N\to\infty}\tilde{\omega}_{k}=\omega_{k}=\frac{2\pi k}{\beta\hbar} \ , \ |k|<\bar{M},
\end{equation}
where $\bar{M}=(M-1)/2$.
In this limit, the $M$ lowest modes (the Matsubara modes) become
Fourier coefficients of the position $q(\tau)=q_l$ (with $\tau=\beta_N\hbar l$), which means that $q(\tau)$ can be built from a superposition of Matsubara modes as
\begin{equation}
q(\tau)=Q_{0}+\sqrt{2}\sum_{k=1}^{\bar{M}}\sin(\omega_{k}\tau)Q_{k}+\cos(\omega_{k}\tau)Q_{-k}.
\end{equation}
The significance of working with the Matsubara modes is that $q(\tau)$ is a smooth and continuous function of the imaginary time variable $\tau$.\cite{Hele2015Mats} This will
not in general be true if $q(\tau)$ is built of both Matsubara and non-Matsubara modes, where the later will give rise to non-smooth  (non-Boltzmann) distributions. Note that quite remarkably, at time $t=0$, one can integrate out the non-Matsubara modes of Eq. (\ref{eq:Kubo_NM}) giving rise to an alternative expression for the Kubo transform in the limit $M\to\infty$, $M << N$\cite{Willatt2018,Ceperley1995,Chakravarty1997,Chakravarty1998,Freeman1984}
\begin{equation}
\label{eq:Matsubara_zero_time}
K_{AB}^{[M]}(0)=\frac{1}{(2\pi\hbar)^{M}Z_M}\int d\boldsymbol{Q}\int d\boldsymbol{P}
e^{-\beta R_M(\boldsymbol{Q},\boldsymbol{P})}A_M(\boldsymbol{Q})
B_M(\boldsymbol{Q}),
\end{equation} 
where
\begin{eqnarray}
R_M(\boldsymbol{Q},\boldsymbol{P})=\sum_{n=-\bar{M}}^{\bar{M}}
\left(
\frac{P^2_n}{2m}+\frac{m}{2}\omega^{2}_nQ^{2}_n \right)
+ U_M(\boldsymbol{Q}),
\label{eq:RPHMats}
\end{eqnarray}
\begin{eqnarray}
\label{eq:U_M}
U_M(\boldsymbol{Q})=\frac{1}{\beta\hbar}\int_0^{\beta\hbar}{\rm d}\tau \ V(q(\tau)),
\end{eqnarray}
and
\begin{eqnarray}
Z_M=\frac{1}{(2\pi\hbar)^{M}}\int d\boldsymbol{Q}\int d\boldsymbol{P}
e^{-\beta R_M(\boldsymbol{Q},\boldsymbol{P})},
\end{eqnarray}
with $A_M(\boldsymbol{Q})$ and $B_M(\boldsymbol{Q})$ defined analogously to $U_M(\boldsymbol{Q})$. This expression is significant since it implies that only smooth Matsubara modes contribute to the Boltzmann average of the Kubo correlation function at time zero. At finite times, unless the potential is harmonic, non-Matsubara modes will couple to the smooth modes due to Eq. (\ref{eq:Wigner MoyalNM}) and, hence, the distribution would become jagged and detailed balance will not be satisfied. 
It is worth mentioning that the Janus operator (Eq. (\ref{eq:JanusNM})) does not couple different normal modes together. It is only through the sine function in Eq. (\ref{eq:Wigner MoyalNM}) that the Janus operator mixes the modes together in the Liouvillian.

The Matsubara approximation is then to assume that one can neglect the coupling to the non-Matsubara modes for all times and use only the Matsubara modes to describe the time evolution of the system. This is done by neglecting the non-Matsubara mode terms in Eq. (\ref{eq:JanusNM}) which produces the effect of decoupling the non-Matsubara modes from the Matsubara modes in the dynamical evolution. The quantum Liouvillian then reduces to a classical Liouvillian in the Matsubara modes in the limit $N\to\infty$
\begin{equation}
\label{eq:LbarM}
\lim_{N\to\infty}L_{N}\approx
\bar{L}_{M}=\sum_{k=-\bar{M}}^{\bar{M}}\frac{P_{k}}{m}\frac{\partial}{\partial Q_{k}}-\frac{1}{N}\frac{\partial U_{N}(\boldsymbol{Q})}{\partial Q_{k}}\frac{\partial}{\partial P_{k}},
\end{equation}
which happens because in the Matsubara subspace
$\hbar$ is replaced by an effective Planck's constant $\hbar/N$
which in the limit as $N\to\infty$ vanishes. 
The Matsubara approximation to the Kubo transformed correlation function is then
\begin{equation}
K_{AB}^{[M]}(t)=\lim_{N\to\infty}\frac{1}{(2\pi\hbar)^{N}}\int d\boldsymbol{Q}\int d\boldsymbol{P}\left[e^{-\beta\hat{H}}\hat{A}\right]_{\overline{N}}(\boldsymbol{Q},\boldsymbol{P})e^{\bar{L}_{M}t}B(\boldsymbol{Q}).\label{eq:MatsubaraTCFalmost}
\end{equation}
Note that Eq. (\ref{eq:MatsubaraTCFalmost}) still depends on
the non-Matsubara modes through the potential. However, since these modes are decoupled from the Matsubara modes they can be analytically integrated
out to give\cite{Hele2015Mats,footnote_about_delta}
\begin{equation}
K_{AB}^{[M]}(t)=
\frac{1}{2\pi\hbar Z_M}\int d\boldsymbol{Q}\int d\boldsymbol{P}e^{-\beta H_M(\boldsymbol{Q},\boldsymbol{P})}e^{i\beta\theta_M(\boldsymbol{Q},\boldsymbol{P})}A_M(\boldsymbol{Q)}e^{L_{M}t}B_M(\boldsymbol{Q}),\label{eq:MastubaraTCF}
\end{equation}
where it is to be understood that all the variables are now of the Matsubara
modes only. $L_{M}$  is the classical Liouvillian in the Matsubara subspace (Eq. (\ref{eq:LbarM})) but with the Matsubara
potential $U_{M}(\boldsymbol{Q})$ (defined by Eq. (\ref{eq:U_M})) replacing $U_{N}(\boldsymbol{Q})/N$. The Matsubara Hamiltonian is given by
\begin{equation}
H_M(\boldsymbol{Q},\boldsymbol{P})=\sum_{k=-\bar{M}}^{\bar{M}}\frac{P_{k}^{2}}{2m}+U_{M}(\boldsymbol{Q}),
\end{equation}
and the Matsubara phase, which converts what would be a classical Boltzmann distribution into a quantum one, is
\begin{equation}
\theta_M(\boldsymbol{Q},\boldsymbol{P})=\sum_{k=-\bar{M}}^{\bar{M}}\omega_{k}Q_{-k}P_{k}.
\end{equation}

It can be straightforwardly shown that the Matsubara correlation function of Eq. (\ref{eq:MastubaraTCF})
contains all the symmetries of the Kubo transformed correlation function. Furthermore, it can
be shown that the classical dynamics generated by the Matsubara Liouvillian
preserve both the Matsubara phase and the Boltzmann distribution, which ensures the quantum Boltzmann distribution is conserved during the classical evolution of the Matsubara modes. Matsubara dynamics is also exact in the harmonic limit for any correlation function due to the fact that for this particular potential the non-Matsubara modes do not couple to the Matsubara modes. Matsubara dynamics will perform better than Ring Polymer Molecular Dynamics (RPMD)\cite{Craig2004,Habershon2008} and Centroid Molecular Dynamics (CMD)\cite{Cao1994-2,Cao1994-4,Jang1999} in general due to the fact that it explicitly includes the fluctuation dynamics that both CMD and RPMD miss.
This is to date the most general form of classical Boltzmann preserving
dynamics from which it has been shown that RPMD, CMD and the Planetary Model\cite{Smith2015} 
are in fact approximations of Matsubara dynamics\cite{Hele2015,Willatt2018}.

\section{Two-Time Matsubara Dynamics}
\label{sec:Two-Time Matsubara Dynamics}

Having established the key steps of the formulation of single-time Matsubara dynamics, we now present the derivation of the Matsubara approximation for the two-time symmetrized double Kubo transform correlation function. To keep the discussion concise we utilize many of the results from Sec. \ref{sec:Mats}.

\subsection{Path Integral Discretization of the Symmetrized Double Kubo Transform}
\label{subsec:pathintegral}

We start by defining the symmetrized double Kubo transform as \cite{Reichman2000,Jung2018}
\begin{equation}
K_{ABC}^{sym}(t,t')=\frac{1}{Z\beta^{2}}\int_{0}^{\beta}d\lambda\int_{0}^{\beta}d\lambda'\bigl\langle\hat{T}_{\beta} \hat{A}(-i\hbar\lambda)\hat{B}(-i\hbar\lambda'+t)\hat{C}(t')\bigr\rangle,
\label{eq:TwotimeKuboTimeOrdered}
\end{equation}
where $\hat{O}(\tau)=e^{i\hat{H}\tau/\hbar}\hat{O}e^{-i\hat{H}\tau/\hbar}$ and $\hat{T}_{\beta}$ is the imaginary time ordering operator
\begin{eqnarray}
\label{eq:T_hat}
\hat{T}_\beta\hat{O}(-i\hbar\lambda)\hat{P}(-i\hbar\lambda')=
\begin{cases}
\hat{O}(-i\hbar\lambda)\hat{P}(-i\hbar\lambda') \ \ \ \text{if } \lambda>\lambda'\\
\hat{P}(-i\hbar\lambda')\hat{O}(-i\hbar\lambda) \ \ \ \text{if } \lambda<\lambda'
\end{cases}
\end{eqnarray}
Here, $t$ and $t'$ are taken to be independent time variables. 
By using the definition of the imaginary time ordering operator and exchanging the integration limits, Eq. (\ref{eq:TwotimeKuboTimeOrdered}) can be expressed as
\begin{equation}
K_{ABC}^{sym}(t,t') = I_1 + I_2 
\end{equation}
where
\begin{equation}
\label{eq:I_1_init}
I_1 = \frac{1}{Z\beta^{2}}\int_{0}^{\beta}d\lambda\int_{0}^{\lambda}d\lambda'Tr \bigg[e^{-(\beta-\lambda)\hat{H}}\hat{A}e^{-(\lambda-\lambda')\hat{H}}\hat{B}(t)e^{-\lambda'\hat{H}}\hat{C}(t')\bigg]
\end{equation}
and
\begin{equation}
\label{eq:I_2_init}
I_2= \frac{1}{Z\beta^{2}}\int_{0}^{\beta}d\lambda\int_{0}^{\lambda}d\lambda'Tr \bigg[e^{-(\beta-\lambda)\hat{H}}\hat{B}(t)e^{-(\lambda-\lambda')\hat{H}}\hat{A}e^{-\lambda'\hat{H}}\hat{C}(t')\bigg].
\end{equation}
Note
here that $\beta\geq\lambda\geq\lambda'\geq0$, which ensures that there is no backwards imaginary time propagation.
The symmetrized Double Kubo transform is a real function of time that shares the formal properties and symmetries with classical two-time correlation functions (see Appendix of Ref. \citenum{Jung2018})
and has recently been related to the second-order response function of nonlinear spectroscopy.\cite{Jung2018}
 
Following the ideas of the previous section, we discretize the iterated integrals and make the expression more symmetric by inserting $N-1$ identities of the form
\begin{equation}
\hat{1} = e^{i\hat{H}t/\hbar}e^{-i\hat{H}t/\hbar}e^{i\hat{H}t'/\hbar}e^{-i\hat{H}t'/\hbar},
\label{eq:TimeII}
\end{equation}
to obtain
\begin{eqnarray}
\label{eq:I_1}
I_1 & = & \frac{1}{Z_NN^2} \sum^{N}_{k=1} \sum^{k}_{l=1}Tr \bigg[ \bigg(e^{-\beta_N\hat{H}}e^{i\hat{H}t/\hbar}e^{-i\hat{H}t/\hbar}e^{i\hat{H}t'/\hbar}e^{-i\hat{H}t'/\hbar}\bigg)^{N-k-1} \nonumber \\
& & \times e^{-\beta_N\hat{H}} \hat{A}e^{i\hat{H}t/\hbar}e^{-i\hat{H}t/\hbar}e^{i\hat{H}t'/\hbar}e^{-i\hat{H}t'/\hbar}\bigg(e^{-\beta_N\hat{H}}e^{i\hat{H}t/\hbar}e^{-i\hat{H}t/\hbar}e^{i\hat{H}t'/\hbar}e^{-i\hat{H}t'/\hbar}\bigg)^{k-l-1} \nonumber \\
& & \times e^{-\beta_N\hat{H}}e^{i\hat{H}t/\hbar}\hat{B} e^{-i\hat{H}t/\hbar} e^{i\hat{H}t'/\hbar}e^{-i\hat{H}t'/\hbar}\bigg(e^{-\beta_N\hat{H}}e^{i\hat{H}t/\hbar}e^{-i\hat{H}t/\hbar}e^{i\hat{H}t'/\hbar}e^{-i\hat{H}t'/\hbar}\bigg)^{l-1} \nonumber \\
& & \times e^{-\beta_N\hat{H}}e^{i\hat{H}t/\hbar} e^{-i\hat{H}t/\hbar} e^{i\hat{H}t'/\hbar}\hat{C}e^{-i\hat{H}t'/\hbar} \bigg]
\end{eqnarray}
and
\begin{eqnarray}
\label{eq:I_2}
I_2 & = & \frac{1}{Z_NN^2} \sum^{N}_{k=1} \sum^{k}_{l=1}Tr \bigg[ \bigg(e^{-\beta_N\hat{H}}e^{i\hat{H}t/\hbar}e^{-i\hat{H}t/\hbar}e^{i\hat{H}t'/\hbar}e^{-i\hat{H}t'/\hbar}\bigg)^{N-k-1} \nonumber \\
& & \times e^{-\beta_N\hat{H}} e^{i\hat{H}t/\hbar}\hat{B}e^{-i\hat{H}t/\hbar}e^{i\hat{H}t'/\hbar}e^{-i\hat{H}t'/\hbar}\bigg(e^{-\beta_N\hat{H}}e^{i\hat{H}t/\hbar}e^{-i\hat{H}t/\hbar}e^{i\hat{H}t'/\hbar}e^{-i\hat{H}t'/\hbar}\bigg)^{k-l-1} \nonumber \\
& & \times e^{-\beta_N\hat{H}}\hat{A}e^{i\hat{H}t/\hbar} e^{-i\hat{H}t/\hbar} e^{i\hat{H}t'/\hbar}e^{-i\hat{H}t'/\hbar}\bigg(e^{-\beta_N\hat{H}}e^{i\hat{H}t/\hbar}e^{-i\hat{H}t/\hbar}e^{i\hat{H}t'/\hbar}e^{-i\hat{H}t'/\hbar}\bigg)^{l-1}  \nonumber \\
& & \times e^{-\beta_N\hat{H}}e^{i\hat{H}t/\hbar} e^{-i\hat{H}t/\hbar} e^{i\hat{H}t'/\hbar}\hat{C}e^{-i\hat{H}t'/\hbar} \bigg]
. 
\end{eqnarray}
The structure of Eqs. (\ref{eq:I_1}) and (\ref{eq:I_2})
suggests that the interpretation of the double Kubo transforms can be cast in terms of a new block structure of the form $\hat{J}^{(2)}=e^{-\beta\hat{H}}e^{i\hat{H}t/\hbar} e^{-i\hat{H}t/\hbar} e^{i\hat{H}t'/\hbar}e^{-i\hat{H}t'/\hbar}$ with the operators $\hat{A}$ and $\hat{B}$ evaluated in different blocks depending of the sum indexes. Note, however, that in Eq. (\ref{eq:I_1}) the operator $\hat{A}$ is evaluated over all the blocks whereas the operator $\hat{B}$ is evaluated only at particular blocks (constrained by the inner sum index); in Eq. (\ref{eq:I_2}) the opposite behavior is found, i.e. $\hat{B}$ is evaluated over all the blocks whereas $\hat{A}$ is constrained at particular blocks. This is a consequence of the imaginary time ordering found in Eqs. (\ref{eq:I_1_init}) and (\ref{eq:I_2_init}). However, since the underling 
 block structure is the same for both $I_1$ and $I_2$ (i.e. the block structure $\hat{J}^{(2)}$ is common to both traces),    
 the \textit{sum} of $I_1$ and $I_2$ allows the evaluation of both $\hat{A}$ and $\hat{B}$ over all the blocks. This is vital for obtaining a symmetric form of the Kubo transform and will be crucial for constructing the multi-time generalization of Matsubara dynamics (see section \ref{sec:multi-dimensional}).

\begin{figure}[t]
\centering
\begin{subfigure}[l]{0.45\textwidth}
	\centering
	\includegraphics[width=0.75\textwidth]{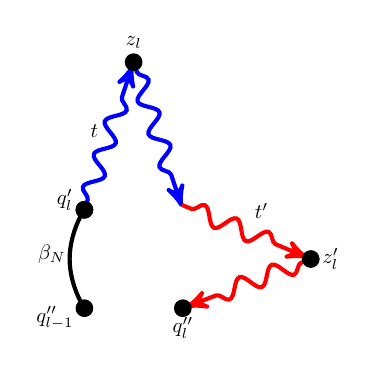} 
\end{subfigure}
\begin{subfigure}[r]{0.45\textwidth}
	\centering
	\includegraphics[width=\textwidth]{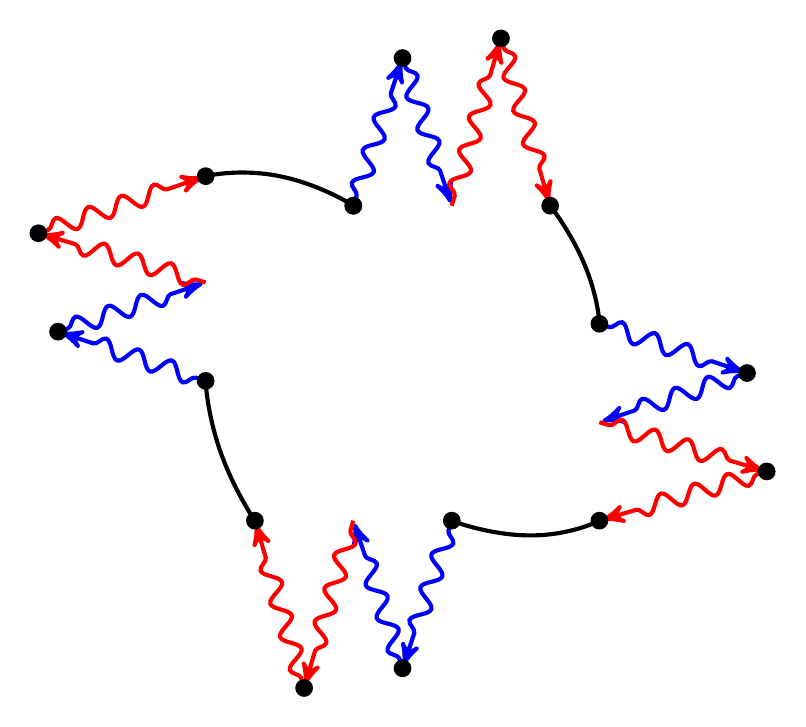}
\end{subfigure}	
\caption{Left: Schematic diagram of the path integral discretization of the  block structure $\hat{J}^{(2)}$ used to obtained the generalized symmetrized double Kubo transform in Eq. (\ref{eq:GenDKubo2}). Blue wavy lines show the $t$ propagation. Red wavy lines show the $t'$ propagation. Black lines show the $\beta_N$ propagation. 
Note that there is no identity inserted between the forward $t$ evolution and the backward $t'$ evolution.
Right: Schematic diagram of the  structure of the generalized symmetrized double Kubo transform time correlation function (Eq. (\ref{eq:GenDKubo2}))
}
\protect\label{fig:2}
\end{figure}

To obtain a path integral representation of the double Kubo transform, we path integral discretize the building blocks $\hat{J}^{(2)}$ in the form (see Fig. \ref{fig:2} for an schematic representation) 
 \begin{eqnarray}
\hat{J}^{(2)}&=&\int dq'_l\int dq''_l\int dz_l\int dz'_l \ 
e^{-\beta_N\hat{H}}\ket{q'_l} \bra{q'_l}
e^{i\hat{H}t/\hbar}\ket{z_l} \bra{z_l}
e^{-i\hat{H}t/\hbar}e^{i\hat{H}t'/\hbar}\ket{z'_l} \bra{z'_l}
e^{-i\hat{H}t'/\hbar}\ket{q''_l} \bra{q''_l}.
\label{eq:building_2}
\end{eqnarray}
and add Eqs. (\ref{eq:I_1}) and (\ref{eq:I_2}) to obtain 
\begin{eqnarray}
\label{eq:DKubopath}
K_{ABC}^{sym,[N]}(t,t')& = & \frac{1}{Z_N} 
\int d\boldsymbol{q'}\int d\boldsymbol{q''}\int d\boldsymbol{z}\int d\boldsymbol{z'} A(\boldsymbol{q'}) B(\boldsymbol{z}) C(\boldsymbol{z'}) \nonumber \\
& & \times \prod_{l=1}^{N}\left\langle q''_{l-1}\right|e^{-\beta_{N}\hat{H}}\left|q'_{l}\right\rangle \left\langle q'_{l}\right|e^{i\hat{H}t/\hbar}\Bigl|z_{l}\Bigr\rangle \nonumber \\
& & \times \Bigl\langle z_{l}\Bigr|e^{-i\hat{H}t/\hbar}e^{i\hat{H}t'/\hbar}\left|z_{l}^{'}\right\rangle \left\langle z_{l}^{'}\right|e^{-i\hat{H}t'/\hbar}\left|q''_{l}\right\rangle.
\label{eq:DKsymmm}
\end{eqnarray}
An schematic diagram of this expression for 
$N=4$ is shown in Fig. \ref{fig:2} and highlights the symmetry of the symmetrized Double Kubo transform.

By making the change of variables from Cartesian coordinates to the 
sum/difference coordinates of Eqs. (\ref{eq:SumDiff1}) and (\ref{eq:SumDiff2}), we can re-express Eq. (\ref{eq:DKubopath}) as
\begin{equation}
\label{eq:GenDKubo2}
K_{ABC}^{sym,[N]}(t,t')=\int d\boldsymbol{q}\int d\boldsymbol{\Delta}\int d\boldsymbol{z}\int d\boldsymbol{z'}A(\boldsymbol{q}+\boldsymbol{\Delta}/2)B(\boldsymbol{z})C(\boldsymbol{z'})\rho(\boldsymbol{q},\boldsymbol{\Delta};\beta)G(\boldsymbol{q},\boldsymbol{\Delta},\boldsymbol{z},\boldsymbol{z'};t,t'),
\end{equation}
where $\rho(\boldsymbol{q},\boldsymbol{\Delta};\beta)$ is defined in Eq. (\ref{eq:rho}) and 
\begin{equation}
G(\boldsymbol{q},\boldsymbol{\ensuremath{\Delta}},\boldsymbol{z},\boldsymbol{z}';t,t')=\prod_{l=1}^{N}\left\langle q_{l}+\frac{\Delta_{l}}{2}\right|e^{i\hat{H}t/\hbar}\Bigl|z_{l}\Bigr\rangle\Bigl\langle z_{l}\Bigr|e^{-i\hat{H}t/\hbar}e^{i\hat{H}t'/\hbar}\left|z_{l}^{'}\right\rangle \left\langle z_{l}^{'}\right|e^{-i\hat{H}t'/\hbar}\left|q_{l}-\frac{\Delta_{l}}{2}\right\rangle .\label{eq:GDKT}
\end{equation}
Eq. (\ref{eq:GenDKubo2}), which we term the Generalized Double Kubo Transformed correlation function, corresponds to an exact path integral representation of the symmetrized double Kubo transform and represents the first important result of this paper. To the best of our knowledge, this is the first time that an expression that emphasizes the symmetry with respect to cyclic permutations of the coordinates of the path integral for the symmetrized double Kubo transform has been presented.

\subsection{Phase space representation}
\label{subsec:phasespace}

A phase space representation of the Generalized Double Kubo Transform Eq. (\ref{eq:GenDKubo2}) can be obtained by  
inserting Dirac delta identities of the form of Eq. (\ref{eq:Dirac_delta})  
to arrive at:
\begin{equation}
K_{ABC}^{sym,[N]}(t,t')=\frac{1}{(2\pi\hbar)^{N}}\int d\boldsymbol{q}\int d\boldsymbol{p}\left[e^{\beta\hat{H}}\hat{A}\right]_{\overline{N}}(\boldsymbol{q},\boldsymbol{p})\left[\hat{B}(t)\hat{C}(t')\right]_{N}(\boldsymbol{q},\boldsymbol{p}),\label{eq:GDKTWigner}
\end{equation}
where $\left[e^{\beta\hat{H}}\hat{A}\right]_{\overline{N}}(\boldsymbol{q},\boldsymbol{p})$ 
has been defined in Eq. (\ref{eq:BadWigner})
and 
\begin{eqnarray}
\left[\hat{B}(t)\hat{C}(t')\right]_{N}(\boldsymbol{q},\boldsymbol{p}) & = & \int d\boldsymbol{\Delta}\int d\boldsymbol{z}\int d\boldsymbol{z'}B(\boldsymbol{z})C(\boldsymbol{z'})\prod_{l=1}^{N}\left\langle q_{l}-\frac{\Delta_{l}}{2}\right|e^{i\hat{H}t/\hbar}\Bigl|z_{l}\Bigr\rangle\nonumber \\
 &  & \times\Bigl\langle z_{l}\Bigr|e^{-i\hat{H}t/\hbar}e^{i\hat{H}t'/\hbar}\left|z_{l}^{'}\right\rangle \left\langle z_{l}^{'}\right|e^{-i\hat{H}t'/\hbar}\left|q_{l}+\frac{\Delta_{l}}{2}\right\rangle e^{i\Delta_{l}p_{l}/\hbar}.
\label{eq:BigW}
\end{eqnarray}
The structure of the two-time Wigner transform in Eq. (\ref{eq:BigW}) involves a double sum over products of one-dimensional Wigner transforms. However, with the use of the Moyal product,\cite{Hillery1984,star_product} which replaces a Wigner transformed product with a product of Wigner transforms,
it can be recast in the more compact form (see Appendix \ref{app:Wigner}):
\begin{equation}
\label{eq:Moyal}
\left[\hat{B}(t)\hat{C}(t')\right]_{N}(\boldsymbol{q},\boldsymbol{p}) = \left[\hat{B}(t)\right]_{N}(\boldsymbol{q},\boldsymbol{p})e^{-i\hbar\hat{\Lambda}_{N}/2}\left[\hat{C}(t')\right]_{N}(\boldsymbol{q},\boldsymbol{p}),
\end{equation}
with $\left[\hat{{\cal O}}(t)\right]_{N}(\boldsymbol{q},\boldsymbol{p})$ defined in Eq. (\ref{eq:GoodWigner}) and the Janus operator $\hat{\Lambda}_{N}$ defined in Eq. (\ref{eq:Janusqp}). Since $\hat{A}$ and $\hat{B}$ only depend on position,
at $t=t'=0$ it follows that:
\begin{equation}
\label{eq:zeroWigner}
\left[\hat{B}(0)\hat{C}(0)\right]_{N}(\boldsymbol{q},\boldsymbol{p}) 
=B(\boldsymbol{q})C(\boldsymbol{q}).
\end{equation}

Noting that due to the identity in Eq. (\ref{eq:Moyal}) and the fact that $t$ and $t'$ are {\it independent} variables it holds that
\begin{equation}
\frac{d^n}{d t^n}\frac{d^m}{d t'^m}\left[\hat{B}(t)\hat{C}(t')\right]_{N}(\boldsymbol{q},\boldsymbol{p}) =
(\hat{L}_N)^n\left[\hat{B}(t)\right]_{N}(\boldsymbol{q},\boldsymbol{p})e^{-i\hbar\hat{\Lambda}_{N}/2}(\hat{L}_N)^m\left[\hat{C}(t')\right]_{N}(\boldsymbol{q},\boldsymbol{p}),
\end{equation}
where $\hat{L}_N$ is the Liouvillian defined in Eq. (\ref{eq:Wigner Moyal}), Eq. (\ref{eq:GDKTWigner}) can be recast in terms of a phase space representation as: 
\begin{equation}
K_{ABC}^{sym,[N]}(t,t')=\frac{1}{(2\pi\hbar)^{N}}\int d\boldsymbol{q}\int d\boldsymbol{p}\left[e^{\beta\hat{H}}\hat{A}\right]_{\overline{N}}(\boldsymbol{q},\boldsymbol{p})e^{\hat{L}_{N}t}B(\boldsymbol{q})e^{-i\hbar\hat{\Lambda}_{N}/2}e^{\hat{L}_{N}t'}C(\boldsymbol{q}).
\label{eq:PhaseSpaceKubosym1}
\end{equation}
Eq. (\ref{eq:PhaseSpaceKubosym1}) corresponds to an exact ring polymer phase average representation of the symmetrized double Kubo transformed correlation and, to the best of our knowledge, represents a novel expression. Note that if not for the Moyal product term, $e^{-i\hbar\hat{\Lambda}_N/2}$, the time evolution of the 
double Kubo transform would involve the action of the (exact) 
Liouvillian operator $\hat{L}_N$ on the observables $\hat{B}$ and $\hat{C}$ for times $t$ and $t'$ independently. The effect of the Moyal term is then to couple trajectories at different times with one another, which is a consequence of the interference nature of quantum mechanics.\cite{Mukamel1996}

\subsection{Two-Time Matsubara Dynamics Approximation}
\label{subsec:Two-Time-Matsubara-Dynamics}

It is straightforward to recast Eq. (\ref{eq:PhaseSpaceKubosym1}) in terms of normal mode coordinates by applying the transformation in Eq. (\ref{eq:NMmatrix}) to obtain 
\begin{eqnarray}
\label{eq:NMKubo}
K_{ABC}^{sym,[N]}(t,t')  & = & \frac{1}{(2\pi\hbar)^{N}Z_N}\int d\boldsymbol{Q}\int d\boldsymbol{P}\left[e^{\beta\hat{H}}\hat{A}\right]_{\overline{N}}(\boldsymbol{Q},\boldsymbol{P})\nonumber \\
 &  & \times e^{\hat{L}_{N}t}B(\boldsymbol{Q})e^{-i\hbar\hat{\Lambda}_{N}/2N}e^{\hat{L}_{N}t'}C(\boldsymbol{Q})
\end{eqnarray}
Note that since in the normal coordinates each derivative with respect to $Q_k$ ($P_k$) brings a factor of $N^{-1/2}$, the Moyal product term in Eq. (\ref{eq:NMKubo}) has an effective Planck constant of $\hbar/N$.

Note that at time $t=t'=0$, where Eq. (\ref{eq:zeroWigner}) holds, the same integration of non-Matsubara modes as in Eq. (\ref{eq:Matsubara_zero_time}) can be done,\cite{Hele2015Mats} resulting in an expression that only involves Matsubara modes. 
Following ideas from the single time Matsubara dynamics one can make the same approximation of decoupling the dynamical evolution of the non-Matsubara modes from the Matsubara modes 
by neglecting the non-Matsubara terms in the Janus operator  
(in the limit as $N\to\infty$ such that $M\ll N$).
Note that this approximation has two consequences in Eq. (\ref{eq:NMKubo}): first, just as in the single-time approximation, the Liouvillian $\hat{L}_N$ is replaced by the classical Matsubara Liouvillian $\bar{L}_{M}$; second, due to the effective Planck constant $\hbar/N$ in the Moyal product, 
$e^{-i\hbar\hat{\Lambda}_{N}/2N}$ is forced to unity in the limit $N\to\infty$. Hence, under the Matsubara approximation, Eq. (\ref{eq:NMKubo}) becomes
\begin{eqnarray}
\label{eq:almostdone}
K_{ABC}^{sym,[N]}(t,t') & = & \lim_{N\to\infty}\frac{1}{(2\pi\hbar)^{N}Z_N}\int d\boldsymbol{Q}\int d\boldsymbol{P}\left[e^{\beta\hat{H}}\hat{A}\right]_{\overline{N}}(\boldsymbol{Q},\boldsymbol{P})\nonumber \\
 &  & \times e^{\bar{L}_{M}t}B(\boldsymbol{Q})e^{\bar{L}_{M}t'}C(\boldsymbol{Q}).
\end{eqnarray}
Note that, just as in the original Matsubara derivation, we have not
used the assumption that $\hbar\to0$ to reach to this classical expression, but rather used the fact that in the Matsubara subspace linearization of the dynamics
arises naturally in the limit as $N\to\infty$ (due to the effective Planck constant $\hbar/N$ going to zero).

Eq. (\ref{eq:almostdone}) still depends on the non-Matsubara modes through the potential term in $\bar{L}_{M}$. However, since the conjugate momenta of the non-Matsubara modes are not present in the Liouvillian nor in $A(\boldsymbol{Q})$, $B(\boldsymbol{Q})$ or $C(\boldsymbol{Q})$ one can integrate out the non-Matsubara modes, just as in the single-time Matsubara formulation, to obtain 
\begin{eqnarray}
K_{ABC}^{sym,[M]}(t,t')  & = & \frac{1}{2\pi\hbar Z_M}\int d\boldsymbol{Q}\int d\boldsymbol{P}e^{-\beta H_M(\boldsymbol{Q},\boldsymbol{P})}e^{i\beta\theta_M(\boldsymbol{Q},\boldsymbol{P})}\nonumber \\
 &  & \times A_M(\boldsymbol{Q})e^{L_{M}t}B_M(\boldsymbol{Q})e^{L_{M}t'}C_M(\boldsymbol{Q}),\label{eq:TTMatsubaraTCF}
\end{eqnarray}
where it is understood that the integration 
is performed over only the Matsubara modes. This expression represents the Matsubara approximation to the symmetrized double Kubo transform and is another major result of the paper. Since the dynamics of both $B(\boldsymbol{Q})$ and $C(\boldsymbol{Q})$
are generated from the classical Matsubara Liouvillian, which conserves both the Matsubara Hamiltonian and phase factor, 
it follows that Eq. (\ref{eq:TTMatsubaraTCF}) satisfies detailed
balance and thus preserves the quantum Boltzmann distribution. Matsubara dynamics is still exact in the harmonic limit for any two-time correlation function and contains all the same symmetries as the exact double Kubo Transform correlation function. In Appendix \ref{app:comp} we present a numerical demonstration of the convergence of Eq. (\ref{eq:TTMatsubaraTCF}) for a model system.

At time $t=t'=0$ one can obtain an alternative expression to Eq. (\ref{eq:TTMatsubaraTCF}). Following Ref. \citenum{Hele2015}, one can perform the same analytic continuation of the phase factor to obtain
\begin{equation}
K_{ABC}^{sym,[M]}(0,0)=\frac{1}{2\pi\hbar Z_M}\int d\boldsymbol{Q}\int d\boldsymbol{P}e^{-\beta R_{M}(\boldsymbol{Q},\boldsymbol{P})}A_M(\boldsymbol{Q})B_M(\boldsymbol{Q})C_M(\boldsymbol{Q}),
\end{equation}
where $R_{M}(\boldsymbol{Q},\boldsymbol{P})$ 
is the ring polymer
Hamiltonian in normal mode coordinates defined in Eq. (\ref{eq:RPHMats}). 
This expression represents an exact (smoothed)
path integral Fourier representation of the statistics of three operators and is equivalent (in the limit of large $M$) to previously derived expressions.\cite{Reichman2000,Jung2018}

\section{Generalization to Multi-time}
\label{sec:multi-dimensional}

Following the derivation of the previous sections, it is possible to perform a generalization of the Matsubara dynamics approximation to 
any order of the symmetrized Kubo transformed multi-time correlation function. In general, the fully symmetrized $n$-th order Kubo transform
is given by
\begin{eqnarray}
K^{sym}(\boldsymbol{t}) \equiv
\frac{1}{Z\beta^{n}}\int_{0}^{\beta}d\lambda_{1}\int_{0}^{\beta}d\lambda_{2}\cdot\cdot\cdot\int_{0}^{\beta}d\lambda_{n-1}\int_{0}^{\beta}d\lambda_{n} \nonumber \\
\times\bigl\langle \hat{T}_{\beta}A_{0}(-i\hbar\lambda_{1})A_{1}(-i\hbar\lambda_{2}+t_{1})\cdot\cdot\cdot A_{n-1}(-i\hbar\lambda_{n}+t_{n-1})A_{n}(t_{n})\bigr\rangle,
\label{eq:FOpsSymKubo}
\end{eqnarray}
where $\boldsymbol{t}=(t_1,t_2,\cdots,t_n)$ and $\hat{T}_{\beta}$ is a generalization of Eq. (\ref{eq:T_hat}). Note that the single Kubo transform (Eq. (\ref{eq:singleKubo})) and the symmetrized Double Kubo transform (Eq. (\ref{eq:TwotimeKuboTimeOrdered})) are special cases of this definition (for $n=1$ and $n=2$, respectively).

By expanding the time ordering operator $\hat{T}_{\beta}$, using properties of the trace and exchanging the integration limits, Eq. (\ref{eq:FOpsSymKubo}) can be expressed as a sum of $n!$ terms of the form
\begin{eqnarray}
K^{sym}(\boldsymbol{t}) = \sum_{j=1}^{n!} I_j(\boldsymbol{t}) 
\label{eq:I_n_init}
\end{eqnarray}
where $I_j(\boldsymbol{t}) $ represents an ordered Kubo transformation with 
$\beta\geq\lambda_1\geq\lambda_2\geq\cdots\lambda_n\geq0$.
An example of what one of these terms looks like is
\begin{eqnarray}
I_1 &=& \frac{1}{Z\beta^{n}}
\int_{0}^{\beta}d\lambda_1
\int_{0}^{\lambda_1}d\lambda_2
\cdots   
\int_{0}^{\lambda_{n-1}}d\lambda_n   \nonumber \\
&&Tr \bigg[
e^{-(\beta-\lambda_1)\hat{H}}\hat{A}_0
e^{-(\lambda_1-\lambda_2)\hat{H}}\hat{A}_1(t_1)
\cdots
e^{-\lambda_n\hat{H}}\hat{A}_n(t_n)
\bigg].
\end{eqnarray}
Note that $n!$ accounts for all the possible permutations of the $n$ operators $\{A_0,A_1,\cdots,A_{n-1}\}$ inside the integral. For the $n=2$ case, the $2!=2$ terms in Eq. (\ref{eq:I_n_init}) are given by Eqs. (\ref{eq:I_1_init}) and (\ref{eq:I_2_init}).

\begin{figure}[t]
\begin{subfigure}[l]{0.45\textwidth}
	\centering
	\includegraphics[width=0.75\textwidth]{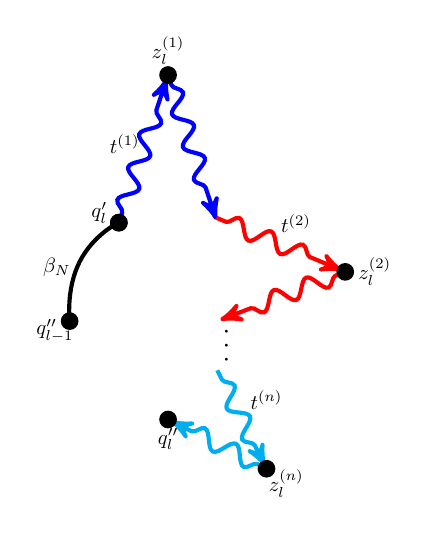} 
\end{subfigure}
\begin{subfigure}[r]{0.45\textwidth}
	\centering
	\includegraphics[width=\textwidth]{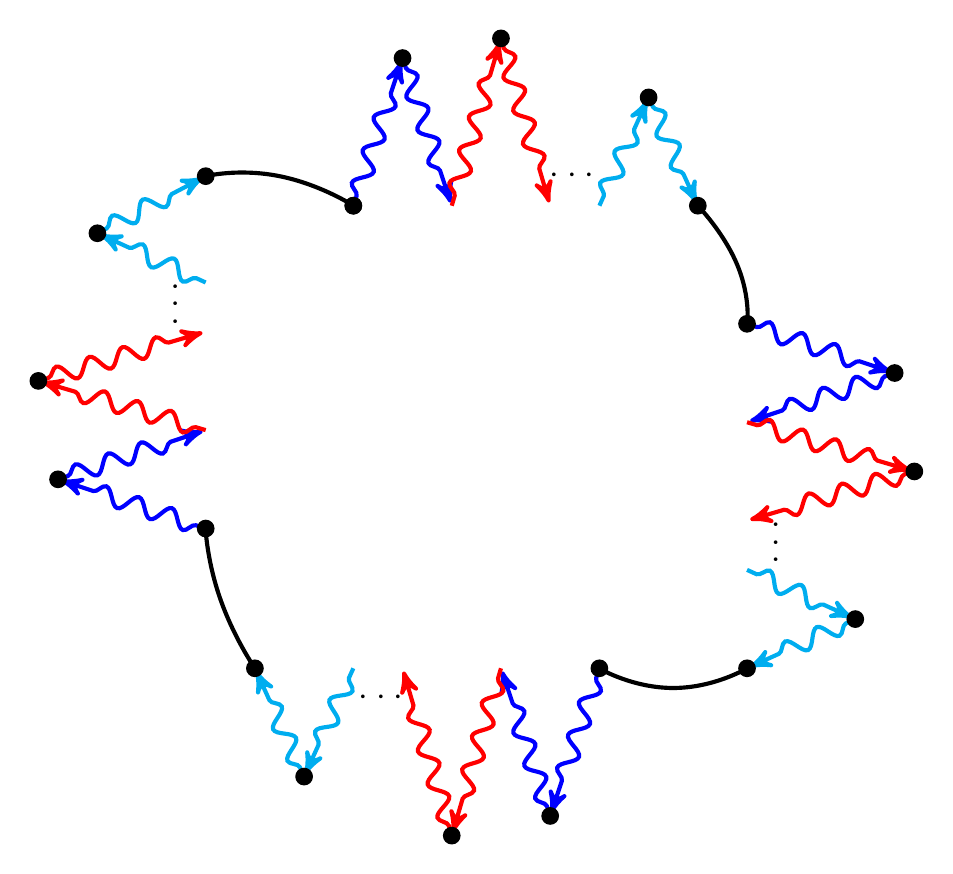}
\end{subfigure}	
\caption{Left: Schematic diagram of the path integral discretization of the block structure $\hat{J}^{(n)}$ (Eq. (\ref{eq:Jn})) used to obtained Eq. (\ref{eq:MTGENK}). Right: Schematic diagram of the  structure of  the general multi-time symmetrized Kubo transform time correlation function (Eq. (\ref{eq:MTGENK}))
 for $N=4$.}
\protect\label{fig:3}
\end{figure}

Generalizing ideas from the previous section, one can discretize the integrals and insert $N-1$ identities of the form
\begin{equation}
\hat{1} = 
e^{i\hat{H}t_1/\hbar}e^{-i\hat{H}t_1/\hbar}
e^{i\hat{H}t_2/\hbar}e^{-i\hat{H}t_2/\hbar}
\cdots
e^{i\hat{H}t_n/\hbar}e^{-i\hat{H}t_n/\hbar},
\label{eq:Time_F}
\end{equation}
to obtain for each $I_j$ a symmetric expression that involves repeating blocks of the form
\begin{eqnarray}
\hat{J}^{(n)}=e^{-\beta\hat{H}}
e^{i\hat{H}t_1/\hbar}e^{-i\hat{H}t_1/\hbar}
e^{i\hat{H}t_2/\hbar}e^{-i\hat{H}t_2/\hbar}
\cdots
e^{i\hat{H}t_n/\hbar}e^{-i\hat{H}t_n/\hbar}
\end{eqnarray}
with the operators $\{A_0,A_1,\cdots,A_{n-1}\}$ intercalated in a particular order and evaluated at a particular block. Following the same logic as in the two-time case, and noting that the underling building block structure is the same for all $I_j$ terms, the \textit{sum} of the $n!$ terms allows to evaluate all the $A_n$ operators over all the blocks.  Hence, by path integral discretizing the building blocks in the form (see Fig. \ref{fig:3})
\begin{eqnarray}
\hat{J}^{(n)}&=&\int dq'_l\int dq''_l\int dz^{(1)}_l\int dz^{(2)}_l \cdots\int dz^{(n)}_l \nonumber \\
& &
e^{-\beta_N\hat{H}}\ket{q'_l} \bra{q'_l}
e^{i\hat{H}t_1/\hbar}\ket{z^{(1)}_l}\bra{z^{(1)}_l}
e^{-i\hat{H}t_1/\hbar}e^{i\hat{H}t_2/\hbar}\ket{z^{(2)}_l}
\cdots \nonumber \\
& &
\bra{z^{(n-1)}_l}e^{-i\hat{H}t_{n-1}/\hbar}e^{i\hat{H}t_n/\hbar}\ket{z^{(n)}_l} \bra{z^{(n)}_l}
e^{-i\hat{H}t_n/\hbar} \ket{q''_l} \bra{q''_l}
\label{eq:Jn}
\end{eqnarray}
and performing the change of variables from the 
$\boldsymbol{q'}$ and $\boldsymbol{q''}$ Cartesian coordinates to $\boldsymbol{q}$ and $\boldsymbol{\Delta}$ sum/difference coordinates, the fully symmetrized $n$-th order Kubo transform Eq. (\ref{eq:FOpsSymKubo}) can be cast as
\begin{eqnarray}
\label{eq:GenMKubo}
K^{sym,[N]}(\boldsymbol{t})&=&\int d\boldsymbol{q}\int d\boldsymbol{\Delta}
\int d\boldsymbol{z^{(1)}}\int d\boldsymbol{z^{(2)}} \cdots
\int d\boldsymbol{z^{(n)}} \nonumber \\
& &
A_0(\boldsymbol{q}+\boldsymbol{\Delta}/2)A_1(\boldsymbol{z^{(1)}}) A_2(\boldsymbol{z^{(2)}})\cdots A_n(\boldsymbol{z^{(n)}}) 
\rho(\boldsymbol{q},\boldsymbol{\Delta};\beta) \nonumber \\
& & G(\boldsymbol{q},\boldsymbol{\Delta},\boldsymbol{z^{(1)}},\boldsymbol{z^{(2)}},\cdots,\boldsymbol{z^{(n)}};\boldsymbol{t}),
\label{eq:MTGENK}
\end{eqnarray}
where $\rho(\boldsymbol{q},\boldsymbol{\Delta};\beta)$ is defined in Eq. (\ref{eq:rho}) and 
$G(\boldsymbol{q},\boldsymbol{\Delta},\boldsymbol{z},\boldsymbol{z^{(1)}},\cdots,\boldsymbol{z^{(n)}};\boldsymbol{t})$
is a multi-time generalization of Eq. (\ref{eq:GDKT}). Eq. (\ref{eq:MTGENK}) represents the complete multi-time extension of the Generalized symmetrized Kubo transform correlation function
and corresponds to an exact path integral representation of Eq. (\ref{eq:FOpsSymKubo}).
An schematic representation of this expression for $N=4$ is presented in Fig. \ref{fig:3}.

Inserting delta function identities allows  one to 
re-express Eq. (\ref{eq:GenMKubo}) as a phase space average of the form
\begin{eqnarray}
K^{sym,[N]}(\boldsymbol{t}) & = & \frac{1}{(2\pi\hbar)^{N}}\int d\boldsymbol{q}\int d\boldsymbol{p}\left[e^{\beta\hat{H}}\hat{A}_{0}\right]_{\overline{N}}(\boldsymbol{q},\boldsymbol{p})
\left[\hat{A}_{1}(t_1)\hat{A}_{2}(t_2)\cdots\hat{A}_{n}(t_n)\right]_{N}(\boldsymbol{q},\boldsymbol{p}), 
\end{eqnarray}
with $\left[\hat{A}_{1}(t_1)\hat{A}_{2}(t_2)\cdots\hat{A}_{n}(t_n)\right]_{N}(\boldsymbol{q},\boldsymbol{p})$ being the multi-time generalization of Eq. (\ref{eq:BigW}). Using a generalization of the derivation presented in Appendix \ref{app:Wigner}
\begin{eqnarray}
\left[\hat{A}_{1}(t_1)\hat{A}_{2}(t_2)\cdots\hat{A}_{n}(t_n)\right]_{N}&=&
\left[\hat{A}_{1}(t_1)\right]_{N}e^{-i\hbar\hat{\Lambda}_{N}/2}\left[\hat{A}_{2}(t_2)\right]_{N}e^{-i\hbar\hat{\Lambda}_{N}/2}\cdots \nonumber \\
& & 
\left[\hat{A}_{n-1}(t_{n-1})\right]_{N}e^{-i\hbar\hat{\Lambda}_{N}/2}\left[\hat{A}_{n}(t_n)\right]_{N},
\end{eqnarray}
and   
the fact that all the times are independent of one another, 
the $N$ bead phase space
representation of Eq. (\ref{eq:FOpsSymKubo}) is given by
\begin{eqnarray}
\label{eq:phasespaceMKT}
K^{sym,[N]}(\boldsymbol{t}) & = & \frac{1}{(2\pi\hbar)^{N}}\int d\boldsymbol{q}\int d\boldsymbol{p}\left[e^{\beta\hat{H}}\hat{A}_{0}\right]_{\overline{N}}(\boldsymbol{q},\boldsymbol{p})e^{\hat{L}_{N}t_{1}}A_{1}(\boldsymbol{q})e^{-i\hbar\hat{\Lambda}_{N}/2}\nonumber \\
 &  & \times e^{\hat{L}_{N}t_{2}}A_{2}(\boldsymbol{q})\cdot\cdot\cdot e^{\hat{L}_{N}t_{n-1}}A_{n-1}(\boldsymbol{q})e^{-i\hbar\hat{\Lambda}_{N}/2}e^{\hat{L}_{N}t_{n}}A_{n}(\boldsymbol{q}).
\end{eqnarray}

The Matsubara approximation to Eq. (\ref{eq:phasespaceMKT}) can be obtained by transforming to normal modes, taking the limit $N\to\infty$ and neglecting the non-Matsubara modes from the Janus operator $\hat{\Lambda}_N$ to yield
\begin{eqnarray}
K^{sym,[M]}(\boldsymbol{t}) & = & \frac{1}{2\pi\hbar Z_M}\int d\boldsymbol{Q}\int d\boldsymbol{P} \ e^{-\beta H(\boldsymbol{Q},\boldsymbol{P})}e^{i\beta\theta(\boldsymbol{Q},\boldsymbol{P})}\nonumber \\
 &  & \times A_{0}(\boldsymbol{Q})e^{L_{M}t_{1}}A_{1}(\boldsymbol{Q})e^{L_{M}t_{2}}A_{2}(\boldsymbol{Q})\cdots \nonumber \\
 &  & \times e^{L_{M}t_{n-1}}A_{n-1}(\boldsymbol{Q})e^{L_{M}t_{n}}A_{n}(\boldsymbol{Q}).
 \label{eq:MTMats}
\end{eqnarray}
The multi-time generalization of Matsubara dynamics is, just as the single-time counterpart, a classical-like approximation to the Kubo transformed quantum multi-time correlation function that preserves the Boltzmann distribution. Eq. (\ref{eq:MTMats}) is the most general form of Matsubara dynamics under the Born-Oppenheimer approximation from which it can be seen that the single-time formulation is just a special case of it.

\section{Conclusions and Future Work}
\label{sec:Conc}

In the present work we have presented a multi-time generalization of Matsubara dynamics\cite{Hele2015Mats} for the calculation of multi-time correlation functions. The theory is based on the decoupling of the non-Matsubara modes from the Matsubara modes in the dynamical evolution and provides a general consistent way of
obtaining classical dynamics from quantum dynamics while preserving the quantum Boltzmann statistics.
Any practical approximation that has been developed for the single-time case\cite{Willatt2018,Craig2004,Cao1994-4,Trenins2018} should be straightforwardly applicable.

The multi-time generalization of Matsubara dynamics approximates
multi-time fully symmetrized Kubo transformed correlation functions. The symmetrized nature of these functions guarantees they
are always real-valued. However, for general multi-time correlation functions, other possible Kubo correlation functions exist. For example, we have shown how the second order response function of non-linear spectroscopy can be expressed in terms of symmetrized and asymmetric double Kubo transform, the latter being a purely imaginary function of time.\cite{Jung2018}   
This raises the question if similar Boltzmann conserving classical dynamics approximations to the one presented here can be developed to more general Kubo transformed correlation functions. Future work in this are will be needed to address this question.

Besides deriving the multi-time formulation of Matsubara dynamics, which is the main purpose of the work, the derivation presented here also provides an exact expression for the evaluation of the multi-time symmetrized Kubo transform, both in path integral form (Eq. (\ref{eq:GenMKubo})) or as a phase space average (Eq. (\ref{eq:phasespaceMKT})). Although these expressions are impractical for the calculation of multi-time correlations functions for condensed phases systems, they might serve as starting points for the development of other semi-classical approximations. Work in this direction is currently underway.

\section*{Acknowledgments}

K. A. J. thanks Stuart Althorpe for stimulating discussions during the summer of 2017 that lead to development of this work and for
helpful comments on a early version of the manuscript.
V.S.B. acknowledges support by the Air Force Office of Scientific Research Grant No. FA9550-17-0198 and high-performance computing time from
Yale High Performance Computing Center.
 
\appendix

\section{Proof of Eq. (\ref{eq:Moyal})}
\label{app:Wigner}

To prove that Eq. (\ref{eq:Moyal}) holds, it is useful to re-express Eq. (\ref{eq:BigW}), by noting that $N-1$ of the forward-backward propagations are identities, as
\begin{eqnarray}
\left[\hat{B}(t)\hat{C}(t')\right]_{N}(\boldsymbol{q},\boldsymbol{p})&=&\sum_{j=1}^N \left[\hat{B}(t)\hat{C}(t')\right]_{W}(q_j,p_j) \nonumber \\
&+& \sum_{\substack{j,k=1\\ j\neq k}}^N \left[\hat{B}(t)\right]_{W}(q_j,p_j) \left[\hat{C}(t')\right]_{W}(q_k,p_k) 
\end{eqnarray}
where $\left[\hat{O}\right]_{W}$ defines the Wigner transform of operator $\hat{O}$ as
\begin{eqnarray}
\left[\hat{O}\right]_{W}(q,p)=\int d\Delta e^{ip\Delta/\hbar}\bra{q-\Delta/2}\hat{O}\ket{q+\Delta/2}.
\end{eqnarray}
Noting that the Wigner transform of a product is given by\cite{Hillery1984}
\begin{eqnarray}
\left[\hat{O_1}\hat{O_2}\right]_{W}(q,p)=
\left[\hat{O_1}\right]_{W}(q,p)
e^{-i\hat{\Lambda}\hbar/2}
\left[\hat{O_2}\right]_{W}(q,p)
\end{eqnarray}
where $\hat{\Lambda}$ is the Janus operator (see Eq. \ref{eq:Janusop}),
the terms in the first sum can be further re-expressed as
\begin{eqnarray}
\left[\hat{B}(t)\hat{C}(t')\right]_{N}(\boldsymbol{q},\boldsymbol{p})&=&
\sum_{j=1}^N \left[\hat{B}(t)\right]_{W}(q_j,p_j)
e^{-i\hat{\Lambda}_j\hbar/2}
\left[\hat{C}(t')\right]_{W}(q_j,p_j) \nonumber \\
&+& \sum_{\substack{j,k=1\\ j\neq k}}^N \left[\hat{B}(t)\right]_{W}(q_j,p_j) \left[\hat{C}(t')\right]_{W}(q_k,p_k). 
\end{eqnarray}

Using the definition of the Janus operator $\hat{\Lambda}_j$ and noting that the leading order of an exponential is 1, the previous expression can be cast as
\begin{eqnarray}
\left[\hat{B}(t)\hat{C}(t')\right]_{N}(\boldsymbol{q},\boldsymbol{p})&=&
\sum_{j,k=1}^N \left[\hat{B}(t)\right]_{W}(q_j,p_j)
e^{-i\hat{\Lambda}_j\hbar/2}
\left[\hat{C}(t')\right]_{W}(q_k,p_k).
\end{eqnarray}

Noting that mixed derivatives inside the sum are zero, namely
\begin{eqnarray}
\left[\hat{B}(t)\right]_{W}(q_j,p_j)
e^{-i\hat{\Lambda}_j\hbar/2}=
\left[\hat{B}(t)\right]_{W}(q_j,p_j)
e^{-i\sum_{l=1}^{N}\hat{\Lambda}_l\hbar/2}
\end{eqnarray}
 the previous equation can be recast as
\begin{eqnarray}
\left[\hat{B}(t)\hat{C}(t')\right]_{N}(\boldsymbol{q},\boldsymbol{p})&=&
\left(\sum_{j=1}^N \left[\hat{B}(t)\right]_{W}(q_j,p_j)\right)
e^{-i\sum_{l=1}^{N}\hat{\Lambda}_l\hbar/2}
\left(\sum_{k=1}^N
\left[\hat{C}(t')\right]_{W}(q_k,p_k)\right)
\end{eqnarray} 

Recognizing that $\sum_{l=1}^{N}\hat{\Lambda}_l=
\hat{\Lambda}_N$ and that 
\begin{eqnarray}
\sum_{j=1}^N \left[\hat{B}(t)\right]_{W}(q_j,p_j)=\left[\hat{B}(t)\right]_{N}(\boldsymbol{q},\boldsymbol{p})
\end{eqnarray}
it follows that
\begin{eqnarray}
\left[\hat{B}(t)\hat{C}(t')\right]_{N}(\boldsymbol{q},\boldsymbol{p})&=&
\left[\hat{B}(t)\right]_{N}(\boldsymbol{q},\boldsymbol{p})
e^{-i\hat{\Lambda}_N\hbar/2}
\left[\hat{C}(t')\right]_{N}(\boldsymbol{q},\boldsymbol{p})
\end{eqnarray}

\section{Matsubara Dynamics Computational Details}
\label{app:comp}

\begin{figure}[t]
\centering
\includegraphics[width=0.8\textwidth]{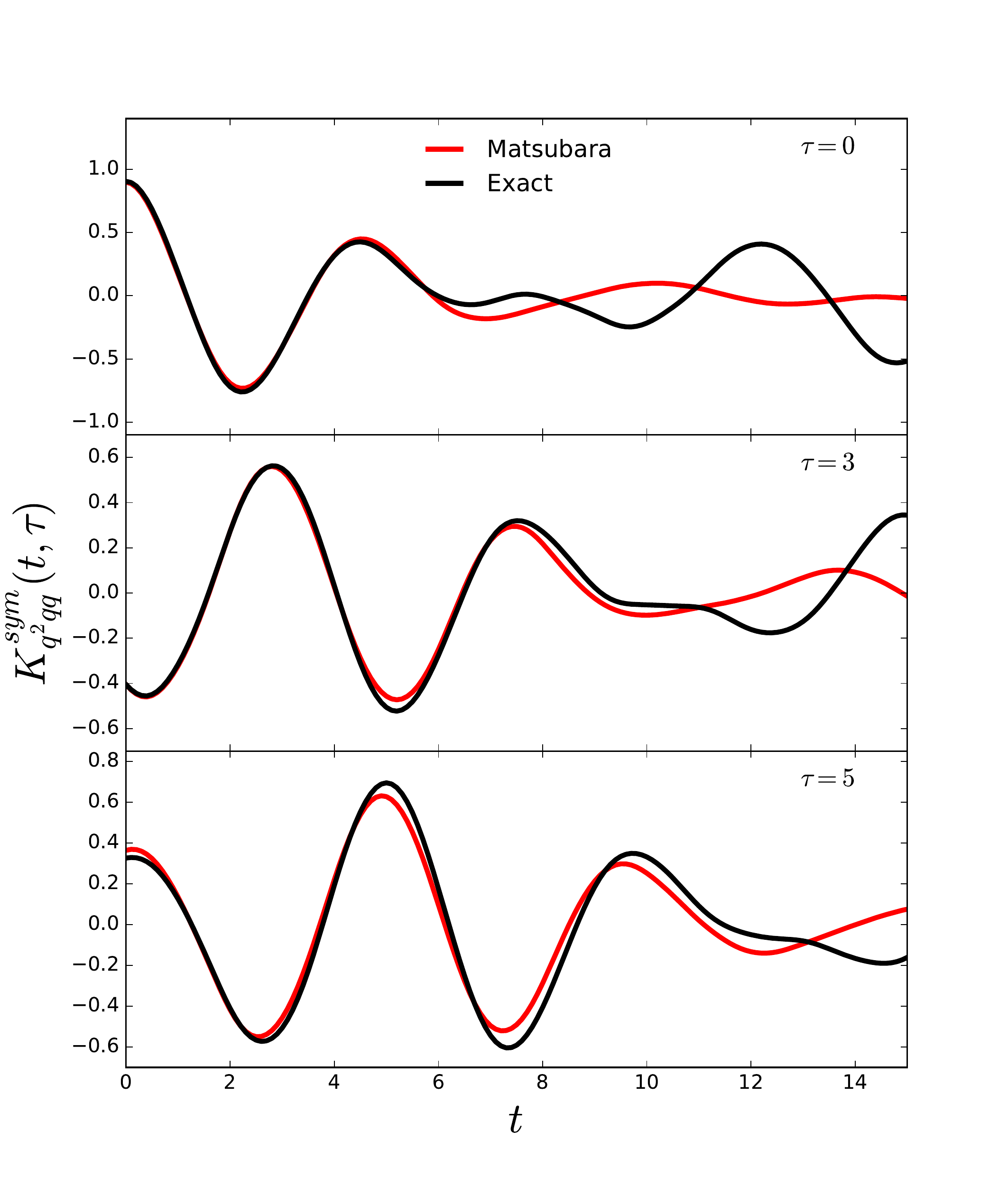} 
\caption{Comparison of the two-time Matsubara approximation and the exact correlation function.}
\protect\label{fig:4}
\end{figure}

To test the performance of the two-time Matsubara dynamics, we perform numerical comparisons between Eq. (\ref{eq:TTMatsubaraTCF}) and the exact result for a model potential. We considered the quartic potential $V(q)=\frac{1}{4}q^4$ 
and evaluate the correlation $K^{sym}_{q^2qq}(t,t')$ for a temperature $\beta=1$ (atomic units are used). Note that this potential represents a severe test for any method that 
neglects (real-time) quantum phase information.
In Fig. \ref{fig:4} we present comparisons between exact results and the Matsubara dynamics for three different cuts along the $t'$ axis.
At short times, Matsubara dynamics it is seen to be an excellent approximation, but the accuracy decreases as either $t$ or $t'$ increases. Note that at time $t=t'=0$ Matsubara dynamics is exact. The Matsubara correlation function was calculated entirely in the normal mode representation for polynomial potentials as described in the supplemental information of Ref. \citenum{Hele2015Mats}. The initial normal mode position and momenta were sampled from a normal distribution and thermalized using the Metropolis-Hastings algorithm by sampling from the Matsubara Hamiltonian. Approximately $1\times10^{7}$ thermalization steps were needed. A long molecular dynamics simulation was then run on the thermalized configuration with a time step of $0.1$ a.u. with momentum resampling every 15 a.u. from the Hamiltonian distribution for each mode to compute the correlation function
\begin{equation}
K_{ABC}^{sym,[M]}(t,t') = \frac{\langle e^{i\beta\theta_M(\boldsymbol{Q},\boldsymbol{P})}A_M(\boldsymbol{Q})B_M(\boldsymbol{Q}_t)C_M(\boldsymbol{Q}_{t'}) \rangle}{\langle e^{i\beta\theta_M(\boldsymbol{Q},\boldsymbol{P})}\rangle},
\end{equation}
where
\begin{equation}
\langle \cdots \rangle = \frac{\int d\boldsymbol{Q}\int d\boldsymbol{P} e^{-\beta H_M(\boldsymbol{Q},\boldsymbol{P})}  \cdots}{\int d\boldsymbol{Q}\int d\boldsymbol{P} e^{-\beta H_M(\boldsymbol{Q},\boldsymbol{P})}}.
\end{equation}
We found that $M=5$ Matsubara modes were sufficient for reaching convergence. The convergence of the Matsubara two-time correlation function is slower than that of the single-time and adding more Matsubara modes requires much larger samples restricting the calculation to the high temperature regime.


\bibliography{biblo}

\begin{thebibliography}{10}
\bibitem{footnote_about_delta}
In the Matsubara limit the $\Delta$ depedence is integrated out leaving the
  observable $A$ solely as a function of $\boldsymbol{Q}$.
\bibitem{star_product}
This is also known as the phase-space star product.
\end{thebibliography}
\bibliographystyle{biblo}

\end{document}